\documentclass[intlimits,twoside,a4paper]{article}

\usepackage[cp1251]{inputenc}

\usepackage{amsmath,amssymb}
\usepackage{graphicx}

\usepackage[eqsecnum]{cmpj3}

\usepackage{bm}

\issue{2019}{22}{4}{43604}
\doinumber{10.5488/CMP.22.43604}
\title{Impact of elastic heterogeneity on the propagation of vibrations at finite temperatures in glasses}
\author[H. Mizuno, S. Mossa]{H. Mizuno\refaddr{label1}, S. Mossa\refaddr{label2,label3}}
\addresses{
\addr{label1} Graduate School of Arts and Sciences, The University of Tokyo, 153-8902 Tokyo, Japan
\addr{label2} University Grenoble Alpes, CEA, CNRS, IRIG, SyMMES, F-38000 Grenoble, France
\addr{label3} Institut Laue-Langevin, BP 156, F-38042 Grenoble Cedex 9, France
}

\date{Received	June 19, 2019}
\begin{document}
\maketitle
\begin{abstract}
Some aspects of how sound waves travel through disordered solids are still unclear. Recent work has characterized a feature of disordered solids which seems to influence vibrational excitations at the mesoscales, local elastic heterogeneity. Sound waves propagation has been demonstrated to be strongly affected by inhomogeneous mechanical features of the materials which add to the standard anharmonic couplings, amounting to extremely complex transport properties at finite temperatures. Here, we address these issues for the case of a simple atomic glass former, by Molecular Dynamics computer simulation. In particular, we focus on the transverse components of the vibrational excitations in terms of dynamic structure factors, and characterize the temperature dependence of sound dispersion and attenuation in an extended frequency range. We provide a complete picture of how elastic heterogeneity determines transport of vibrational excitations, also based on a direct comparison of the numerical data with the predictions of the heterogeneous elastic theory.
\pacs 63.50.+x, 43.20.+g
\end{abstract}
\section{Introduction} 
\label{sec:introduction}
The notion of phonon~\cite{kettel,Ashcroft} together with the Boltzmann transport equation~\cite{McGaughey,Broido_2007,Turney2_2009} efficiently describes vibrational excitations in crystals, ranging from the harmonic approximation to temperatures where anharmonic couplings induce finite life-times of the former. (The Akhiezer effect~\cite{Akhieser_1939,Maris_1971}, which is strictly valid for low-frequency sound modes and produces a $\propto \Omega^2 T$ dependence for the anharmonic sound attenuation, will be mentioned below.) The situation is comprehensively more complex for disordered solids~\cite{lowtem,zeller1971thermal,pohl2002low,lubchenko2007microscopic,klinger2010soft}, in particular glasses, where the anharmonic effects coexist with those due to the structural disorder. In these conditions, the phonon-based picture is recovered in the continuum limit, at small wave-vectors (frequencies). By contrast, when mechanisms taking place at (nano-)scales comparable to the typical atomic distances are relevant, more sophisticated concepts are needed. 

Experimental evidences point to anomalous acoustic excitations in glasses in the THz-GHz regime, including sound softening~\cite{monaco_2009} and Rayleigh-like strong scattering~\cite{masciovecchio2006evidence,devos2008hypersound,baldi2010sound}. Among many other works, our~\cite{monaco2009anomalous} has also provided insight on these matters, by numerical simulations. For instance, we highlighted a clear crossover in the apparent inverse life-time of acoustic-like excitations, $1/\tau=\Gamma$, from a Rayleigh-like scattering, $\Gamma\propto\Omega^4$, to a disorder-induced broadening, $\Gamma\propto\Omega^2$, moving from low to high frequency. We also found that the crossover frequency, $\Omega_\text{co}$, is not far from the Ioffe-Regel frequency, $\Omega_\text{IR}$, implying that sound waves start to loose their plane wave character at $\Omega_\text{co}$. In addition, in the same $\Omega$-regime, peculiar features such as excess vibrational states, the so-called Boson Peak (BP)~\cite{buchenau_1984,Malinovsky_1991}, were observed. Since sound waves in glasses are described as superpositions of different vibrational modes~\cite{taraskin_2000}, these features are strongly related to the anomalous sound waves propagation. As an example, an universal connection between transverse sound waves and the BP has been proposed~\cite{monaco2009anomalous,Shintani_2008}. Rationalizing these observations still presents problematic aspects. One can, for instance, reasonably expect that vibrations interact with a spatially varying field associated to the disorder, undergoing substantial damping which adds to that associated to anharmonic couplings. On the one hand, this poses the problem of establishing the relative impact of the two mechanisms on the departure from the macroscopic limit, on approaching the nanoscale. On the other hand, it forces to identify the true nature of the interacting field.

Recently, evidences have accumulated (including some of our own works) which point to a mechanical origin of sound damping, in the form of substantial {\em elastic heterogeneities}~\cite{Duval_1998}. Simulation~\cite{yoshimoto_2004,tsamados_2009,makke_2011} and experimental~\cite{Wagner_2011,Hufnagel_2015} works have indeed demonstrated that disordered solids exhibit inhomogeneous mechanical response at the nano-scale, i.e., elastic moduli do not simply assume the hydrodynamic (continuum) values but rather they fluctuate around the hydrodynamic values, with a non-negligible distribution width. These elastic heterogeneities generate in turn {\em non-affine} deformations~\cite{DiDonna_2005,Maloney_2006}, which add to the applied affine deformation field inducing a significant reduction in elastic moduli~\cite{tanguy_2002,Zaccone_2011}. Following the non-affine deformation, particles turn out to be displaced in a correlated manner, characterized by a typical mesoscopic correlation length-scale~\cite{leonforte_2005,leonforte_2006}. It is natural to expect that interaction with this non-affine displacement amounts to profound modifications of the sound waves transport. In recent simulation works~\cite{Mizuno2_2013,Mizuno_2014,Mizuno_2016}, we have provided strong evidences of this direct correlation between sound waves features and the heterogeneous mechanical properties. Based on these ideas, W. Schirmacher and co-workers~\cite{schirmacher_2006,schirmacher_2007,Schirmacher_2015} have developed a heterogeneous elasticity theory (HET), capable of reproducing numerous of the above features, including sound softening, Rayleigh-like scattering, and the Boson peak. 

Differently to recent insightful theoretical works~\cite{Wyart_2010,DeGiuli_2014,Gelin_2016,Bouchbinder_2018,angelani2018probing,Mizuno_2018,Tong_2019,Wang_2018,Moriel_2019} which have focused on disordered systems at zero temperature, therefore disregarding the effect of anharmonicities, here we investigate the sound properties of a standard atomic glass, at variable, finite temperatures. In particular, we have determined the transverse component of the dynamic structure factor, by simulating extremely large glassy samples, and provided sound dispersions and attenuation rates in an extended frequency range. Building on our previous work, in particular~\cite{mizuno2019unpub}, and adopting the heterogeneous elastic theory as a guide-line for the discussion, we highlight clearly and in a very direct way, the temperature dependent correlation between the sound transport features and the elastic heterogeneity modifications.
\section{Methods} 
\label{sec:methods}
{\bf The model and simulation details.} We have studied by Molecular Dynamics (MD) simulations glassy systems formed by $N$ mono-dispersed point particles, of mass $m$ and diameter $\sigma$, interacting via a pairwise Lennard-Jones (LJ) potential, $V(r)=4\epsilon[(\sigma/r)^{12}-(\sigma/r)^6]$, with $r=r_{ij}$ the distance between particles $i$ and $j$. The potential $V(r)$ is cut-off and shifted at $r_\text c=2.5\sigma$.~\footnote{The discontinuity of the derivative $V'(r)$ at $r_\text c$ can, in general, modify some harmonic vibrational features at $T=0$~\cite{Shimada_2017}. Based on previous works~\cite{monaco2009anomalous,Mizuno3_2015}, however, we do not expect notable modifications in sound wave propagation at finite temperatures.} $m$, $\sigma$, $\epsilon$, and  $\tau=(m \sigma^2 / \epsilon)^{1/2}$ are the units of mass, length, energy, and time, respectively, while temperature is in units of $\epsilon/k_\text{B}$ ($k_\text{B}$ is the Boltzmann constant). If we consider the case of argon as a reference, temperature is in units of $\epsilon / k_\text{B}=125.2$~K, while length and time scales are expressed in units of $\sigma=3.405$~\AA\, and $\tau=2.11$~ps. We have used periodic boundary conditions in cubic simulation boxes of size $L$, such that the number density is $\hat{\rho}=N/L^3=1.015$. At $\hat{\rho}$, the melting and glass-transition temperatures are $T_\text{m} \simeq 1.0$ and $T_\text{g} \simeq 0.4$~\cite{robles2003liquid}, respectively. In order to access the relevant small wave vector ($q$) region relevant here, we have considered $N=4000$ to $1000188$, corresponding to $L=15.80$ to $99.51$. In what follows, we show data pertaining to all values of $N$ together, immediately verifying the absence of any finite size effects.

Initialization runs were conducted in the micro-canonical ($NVE$) ensemble at the temperature $T=2.0$ in the liquid phase, followed by a fast quench with a rate $\rd T/\rd t\approx 400$ down to $T = 10^{-3}$, well below the glass-transition temperature $T_\text{g}$. Next, the systems were heated to $T=10^{-2}$, $10^{-1}$, $2\times 10^{-1}$, still below $T_\text{g}$. At all values of $T$, the quenched glass samples were relaxed for a time (dependent on $N$) sufficient to eliminate any drift in the total energy, and recover energy fluctuations conforming to the equipartition theorem, with a specific heat per particle $c_v\simeq 3$. Following thermalization, we performed the production runs for a ($N$-dependent) total time sufficient to obtain the desired $\omega$-resolution, always well below the smallest calculated line width. We have used the velocity Verlet algorithm for the integration of the equations of motion, with a time step $\delta t=5\times 10^{-3}$. We emphasize that no aging effects were observed in the time evolution of total energy throughout the production runs. All simulations have been realized with the large-scale massively parallel MD simulation tool LAMMPS~\cite{plimpton1995fast}.

{\bf The dynamical structure factor.} Sound waves propagation has been investigated in terms of the transverse component of the dynamical structure factors~\cite{monaco2009anomalous,Mizuno_2014} at wave vector $\mathbf{q}$ and frequency $\omega$,
\begin{equation} 
\label{eq:stqomega}
S_\text{T}(q,\omega) = \frac{1}{2 \piup N} \left( \frac{q}{\omega} \right)^2 \int \rd t \big\langle \mathbf{j}_\text{T}(q,t)\cdot\mathbf{j}^{\dagger}_\text{T}(q,0) \big\rangle e^{\ri \omega t},
\end{equation}
with the transverse current vector $\mathbf{j}_\text{T}(q,t) = \sum_{i=1}^{N} \left\{ \mathbf{v}_i(t) - \left[ \mathbf{v}_i(t) \cdot \widehat{\mathbf{q}} \right] \widehat{\mathbf{q}} \right\} \exp\{\ri \mathbf{q} \cdot \mathbf{r}_i(t) \}$. Here, $q=\vert \mathbf{q} \vert$, $\hat{\mathbf{q}} = \mathbf{q}/q$, and $\langle \rangle $ is the thermodynamic average. Note that the simulation box with $N=1000188$ accommodates vibrational modes with $q = 2\piup/L \simeq 0.06$. Since the first diffraction peak in the static structure factor, $S(q)$, is at $q_m \simeq 7$ (corresponding to an average nearest-neighbors distance $\simeq 2 \piup /q_m =0.9$), some of our spectra refer to $q$-values $\approx 10^2$ times smaller than the border of the pseudo-Brillouin zone at $q \simeq q_m / 2$.

{\bf The vibrational density of states.} In the discussion we will also consider the vibrational density of state (vDOS), $g(\omega)$~\cite{kettel,Ashcroft}. This was determined by numerically diagonalizing the Hessian matrix (second derivatives of the Hamiltonian), calculated from the coordinates of systems completely relaxed in local potential energy minima (inherent structures). The $g(\omega)$ was then calculated by populating an histogram with the square-root of the obtained (positive) eigenvalues. Similar to $S_\text{T}(q,\omega)$, we have considered system sizes up to $N=256000$, in order to adequately sample the $g(\omega)$ on all the relevant $\omega$-range~\cite{monaco2009anomalous}. 

{\bf The spatial distribution of the local shear modulus.} In our discussion we build over the concept of elastic heterogeneity, i.e., the existence of probability distributions of local elastic moduli. In systems like our LJ glass, the bulk ($K$) and shear ($G$) moduli are such that $K\gg G$~\cite{Novikov_2004}, and the shear modulus mostly determines the low-frequency transverse modes behaviour~\cite{Mizuno2_2013,Mizuno_2014,Mizuno_2016}. We, therefore, focus on the probability distributions of the local shear modulus $G^m$ only, which are determined by partitioning the simulation box into an array of cubic domains of linear size $w \simeq 3.16$, identified by an index $m$. We note that a domain includes on average around $30$ particles. For each $m$, $G^m$ was computed by the fluctuation formula~\cite{yoshimoto_2004,lutsko_1988,Wittmer_2013}, where $G^m$ is expressed as $G^m=G_\text A^m-G_\text N^m$. Here, $G_\text A^m$ encodes the elastic response to {\em affine} deformations, where the particles follow the applied affine strain field. By contrast, the {\em non-affine} modulus, $G_\text N^m$, which contributes negatively to the overall modulus, includes the effects associated to additional particle displacements that deviate from the applied affine field. We have followed the same procedure indicated as the ``fully local approach'' in~\cite{Mizuno_2013}. We add two observations. First, since we have demonstrated that the $G^m$ calculation is insensitive to finite system size effects~\cite{Mizuno_2013}, we have used $N=4000$ ($L=15.80$). Second, in the calculations we have shifted potential {\em and} its first derivative, so that both potential and forces are continuous at $r=r_\text c$. Indeed, in~\cite{Mizuno3_2015} we demonstrated that the local modulus values are modified by non-linearities associated to discontinuities in the forces at $r_\text c$. 
%
\begin{figure}[!t]
\centering
\includegraphics[width=0.6\textwidth]{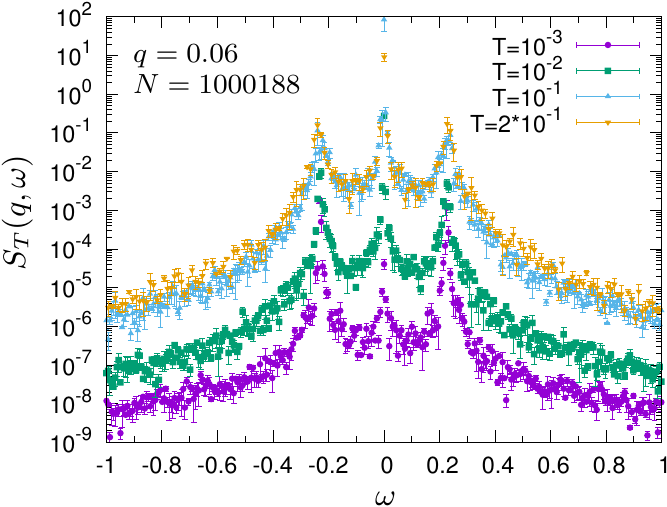}
\caption{(Colour online) Transverse dynamical structure factors, $S_\text{T}(q,\omega)$, for the investigated LJ glass with $N = 1000188$ atoms, at the lowest accessible wave number $q \simeq 0.06$. The different symbols correspond to the indicated temperatures, all below the glass-transition temperature $T_\text{g}\simeq 0.4$.}
\label{fig:spectra}
\end{figure}
\vspace{-4mm}
\section{Results} 
\label{sec:results}
\subsection{Characterizing the transverse dynamical structure factor} 
\label{subsect:transverse}
{\bf The transverse dynamical structure factor.} In figure~\ref{fig:spectra}, we show a selected set of data for $S_\text{T}(q,\omega)$, calculated from equation~(\ref{eq:stqomega}) for $N = 1000188$, at the lowest wave number, $q=0.06$. The indicated values of $T$, all below $T_\text{g} \simeq 0.4$, increase from top to bottom. The $S_\text{T}(q,\omega)$ spectra are characterized by two symmetric Brillouin peaks, in addition to the elastic feature at $\omega=0$. As $T$ increases, the Brillouin peaks move towards higher frequencies, with an increasing total intensity and broadening, as expected. We can extract quantitative information from these data by the customary procedure of fitting the points in the spectral region around the Brillouin peaks to the damped harmonic oscillator (DHO) model~\cite{sette1998dynamics},
\begin{equation} 
\label{eq:dhom}
I_\text{B}(q,\omega) \propto \frac{\Gamma(q) \Omega^2(q)}{\left[ \omega^2-\Omega^2(q) \right]^2 + \omega^2 \Gamma^2(q)}.
\end{equation}
Here, the parameters $\Omega(q)$ and $\Gamma(q)$ represent the characteristic frequency and inverse life-time (or broadening, full width at half maximum) of the Brillouin excitations, respectively, which we discuss below. In what follows, we will systematically dump the subscript $T$. 

\begin{figure}[!t]
\centering
\includegraphics[width=0.98\textwidth]{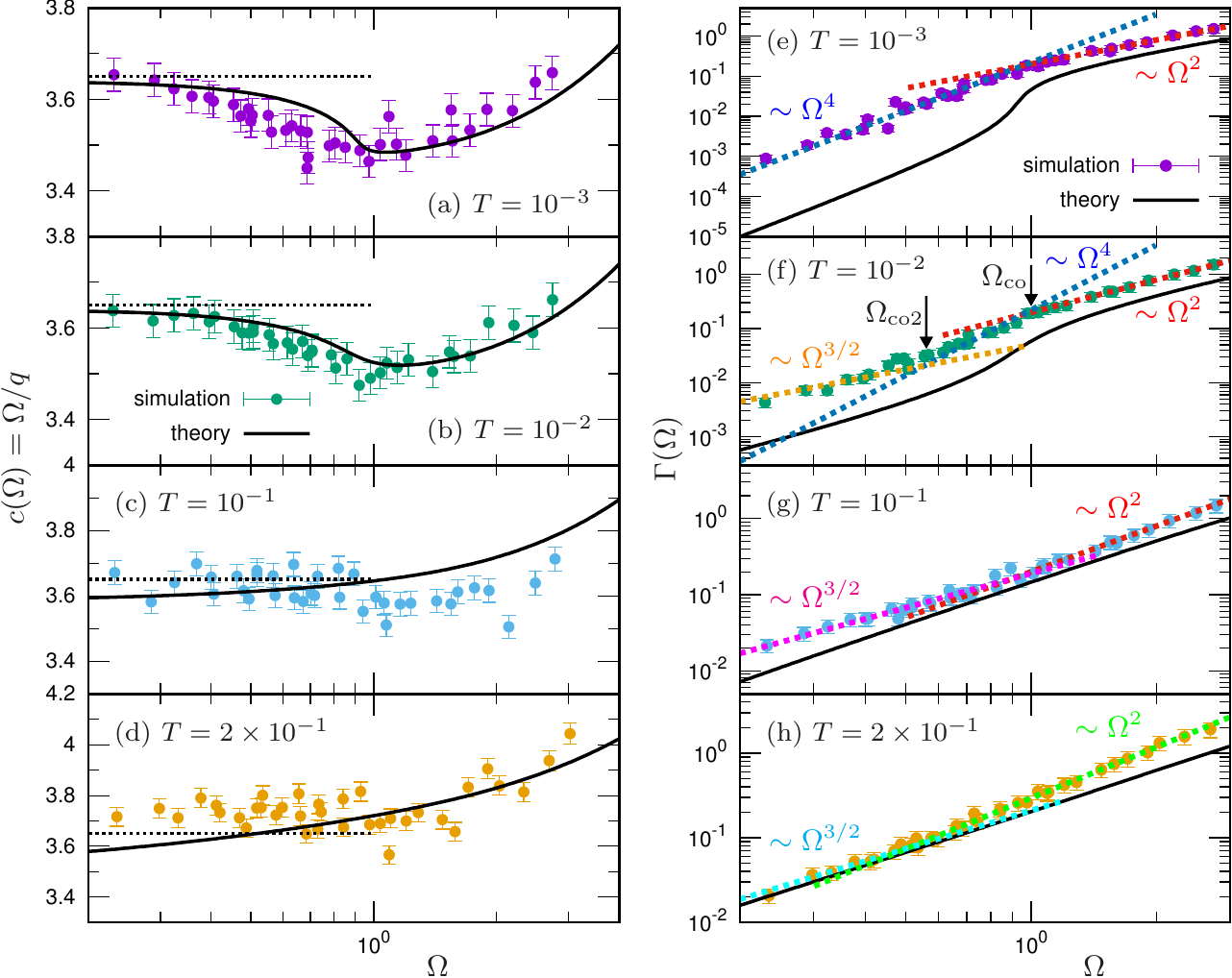}
\caption{(Colour online) (a)--(d): The phase velocity, $c=\Omega(q)/q$, as a function of the corresponding $\Omega$, at the indicated values of $T$. We show the simulation data with symbols and the predictions of HET with solid lines. The horizontal dashed lines indicate the macroscopic limit $v_T=(G/\rho)^{1/2}\simeq 3.65$. (e)--(h): The sound broadening, $\Gamma$, as a function of the corresponding $\Omega$. The dashed lines indicate the power-law scalings valid in the different $\Omega$-regions, with strongly $T$-dependent patterns. We use the same color to indicate power laws which are not modified by changing $T$. The solid lines are the predictions of HET.}
\vspace{-1mm}
\label{fig:gamma}
\end{figure}

{\bf The sound velocity.} From the values of $\Omega(q)$ we can obtain the transverse sound phase velocity, $c(q)= \Omega(q)/q$, shown in figure~\ref{fig:gamma}~(a)--(d) (symbols) as a function of the corresponding $\Omega$, at the indicated values of $T$. In (a) we show the data at the lowest temperature, $T=10^{-3}$, with a velocity increasing with the frequency (the positive dispersion~\cite{ruocco_2000}) for $\Omega > 1$. Next, as already observed in~\cite{monaco2009anomalous}, the velocity shows a minimum before reaching a region of softening, where it increases by decreasing $\Omega$. Eventually, the sound velocity saturates at the correct acoustic value, $c(q \rightarrow 0) \rightarrow v_T = \sqrt{G/\rho} \simeq 3.65$ (dashed line), where $G$ is the shear modulus, and $\rho = m \hat \rho$ is the mass density. The same pattern is still qualitatively present at $T=10^{-2}$ in (b), although the acoustic limit is reached at a slightly higher frequency. At $T=10^{-1}$ in~(c), softening has almost completely disappeared, and a direct crossover from the positive dispersion to the zero frequency limit can be recognized. At the highest temperature, $T=2\times 10^{-1} \simeq T_\text{g}/2$ in (d), the behaviour just discussed is even more evident, with an acoustic limit showing a mild $T$-dependence.

Note that a very similar scenario holds for the longitudinal modes at all values of $T$ (not shown here). Indeed, in reference~\cite{monaco2009anomalous} we have suggested a common origin for both longitudinal and transverse excitations, based on the isotropic elastic medium equation $c_\text L^2 (\omega)\simeq K/\rho+(4/3)c_\text T^2(\omega)$, where we have discarded the frequency dependence of $K(\omega)$, with the replacement of the zero-frequency limit, $K = K(\omega = 0) \simeq 59$. This implies that the knowledge of one of the two velocities allows one to infer the $\omega$-dependence of the other by the above relation. We have checked that this result holds at all values of $T$, and similar observations were also reported in the simulation works of~\cite{Marruzzo_2013,Mizuno_2018}. We note, however, that this is probably not a general conclusion since in~\cite{Mizuno_2014}, for instance, we have demonstrated that, for soft spheres, $K$ does depend on frequency and the above equation is therefore not valid.
%

{\bf The sound broadening.} Our results for the transverse sound broadening, of unprecedented quality, are shown in figure~\ref{fig:gamma} (e)--(h), where we plot $\Gamma$ as extracted from equation~(\ref{eq:dhom}), as a function of the corresponding $\Omega$. The frequency {\em and} temperature dependences of these data are very complex~\cite{mizuno2019unpub}. At the lowest $T=10^{-3}$ in (e), a clear crossover occurs between the high-frequency disorder-controlled behaviour $\propto \Omega^2$~\cite{ruocco1999nondynamic}, and a Rayleigh-like scattering contribution, $\propto\Omega^4$, at lower frequencies~\cite{angelani2000frustration}. As already noticed, the crossover frequency $\Omega_\text{co} \simeq 1$, is not far from the calculated BP frequency, $\omega_\text{BP} \simeq 2$, which is compatible with the Ioffe-Regel frequency, $\Omega_\text{IR} \simeq \omega_\text{BP}$, as also discussed in~\cite{monaco2009anomalous}~\footnote{In~\cite{Mizuno_2018} we have shown that the crossover occurs at the frequency $\omega_0$ where the vDOS converges to the Debye limit. For the present LJ system, $\omega_0 \simeq 1$~\cite{monaco2009anomalous}, confirming $\Omega_\text{co} \simeq \omega_0$.}. We note that, even at this very low $T$, anharmonic interactions are obviously present and, for instance, still contribute to the thermal conductivity. Their intensity, however, is very low in the investigated frequency range compared to other contributions, while non-negligible effects should be visible at frequencies lower than our available spectral window. By increasing $T$, by contrast, we expect the strength of anharmonicities to increase, by eventually entering our accessible frequency range.

This is indeed what we observe in (f), where at $T=10^{-2}$ we detect a second $T$-dependent crossover, at $\Omega_\text{co2} \simeq 0.6$, between the Rayleigh region and an unexpected low-frequency $\propto\Omega^{3/2}$ regime, reminiscent of the fractal-like attenuation of~\cite{ferrante2013acoustic,marruzzo2013vibrational} (see below). Note that this fractional scaling mechanism is evidently $T$-dependent, whereas the Rayleigh and disorder-controlled regimes are not. Also, by increasing $T$, we expect the two crossover frequencies to eventually merge ($\Omega_\text{co2}\simeq\Omega_\text{co}$) when the strength of the anharmonic couplings becomes comparable to that associated to the effect of the disorder, and the two mechanisms entangle in the entire $\Omega$-range. This is indeed what we observe at $T=10^{-1}$ in (g), where the Rayleigh contribution is taken over by the fractal $\sim \Omega^{3/2}$ behaviour at low frequency. In addition, a quadratic $T$-independent contribution is still visible at high $\Omega$. Finally, at the highest $T=2\times 10^{-1} \simeq T_\text{g}/2$ in panel (h), the intensity is fully $T$-dependent in the entire $\Omega$-range, with an Akhiezer-like $\Omega^2$ scaling prevailing in a large frequency range, while a remnant $\Omega^{3/2}$ behaviour can be only recovered at low $\Omega$. This scenario is similar to that reported recently in the experimental work of reference~\cite{Baldi_2014} for a network glass (sodium silicate), confirming the relevance of our findings for realistic systems.

In~\cite{mizuno2019unpub} we have shown how to rationalize the $\Omega^{3/2}$ fractional scaling described above and predicted in the context of the HET~\cite{marruzzo2013vibrational,ferrante2013acoustic}, by disentangling the effects due, on the one hand, to the anharmonicities, and the elastic heterogeneity on the other hand. Below, we focus on characterizing the local elastic heterogeneity at variable $T$, and linking the latter directly to the spectroscopic parameters in terms of the above theory. 
\begin{figure}[!b]
\centering
\includegraphics[width=0.85\textwidth]{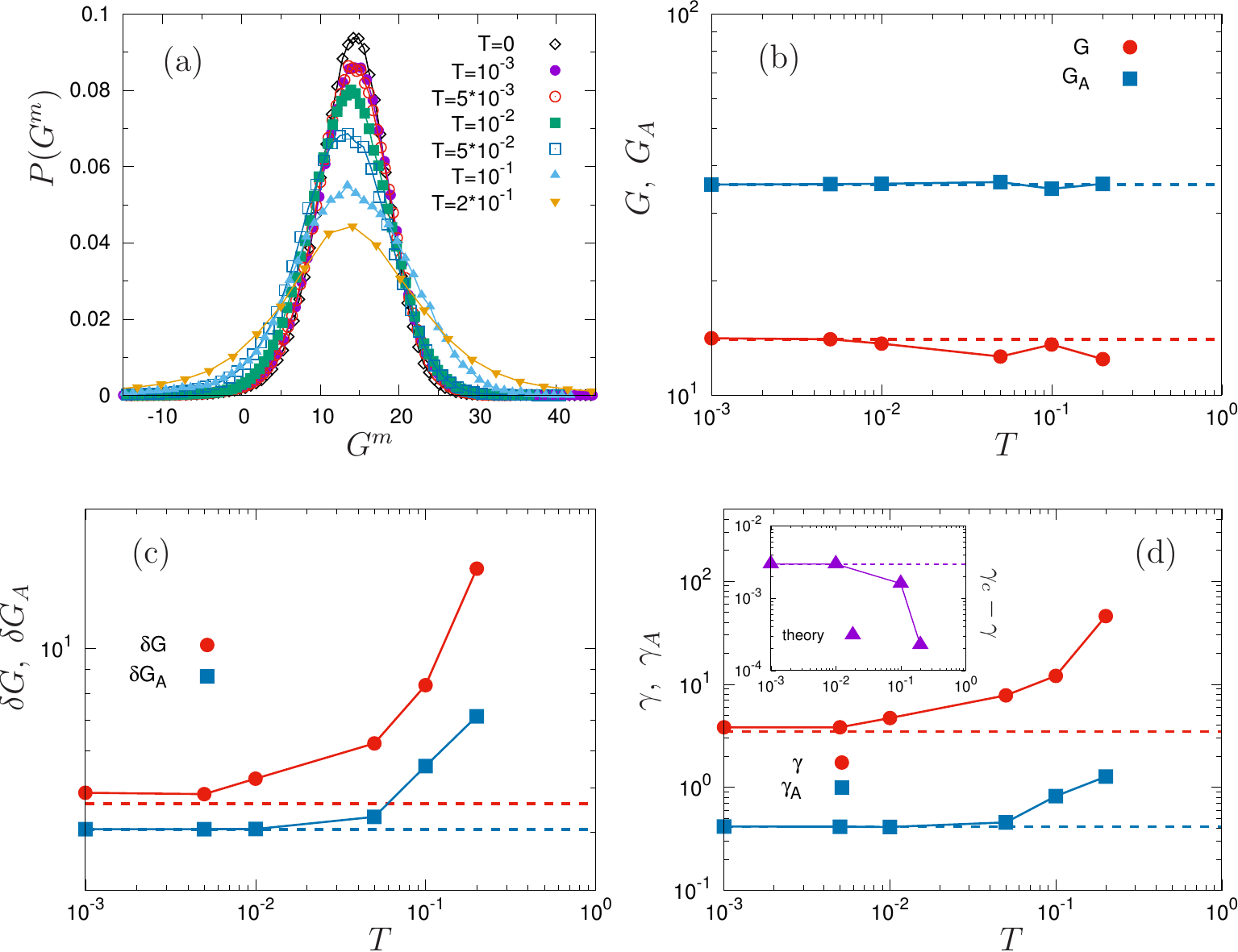}
\caption{(Colour online) (a): Probability distributions, $P(G^m)$, of the local shear modulus, $G^m$, at the indicated values of $T$. (b): The average (macroscopic) value of $P(G^m)$, $G$. (c): The standard deviation of $P(G^m)$, $\delta G$. (d): The parameter $\gamma(T) = \hat{\rho}w^3 \delta G^2/G^2$. In (b)--(d), we also plot $G_\text A$, $\delta G_\text A$, and $\gamma_\text A(T)=\hat{\rho}w^3 \delta G_\text A^2/G_\text A^2$, corresponding to the affine component of the shear modulus, $G_\text A^m$. The horizontal dashed lines indicate the values of the different quantities at $T=0$. In the inset of panel (d), we show the quantity inserted in the HET, $\delta\gamma=\gamma_\text c-\gamma(T)$, with $\gamma_\text c = 0.226$ the critical value where the theory becomes unstable.}
\label{fig:shear}
\end{figure}
\subsection{Uncovering the elastic heterogeneity} 
\label{subsect:elastic}
{\bf The local shear modulus distributions.} We now discuss our numerical determination of the local shear modulus distributions, recalled in section~\ref{sec:methods}. In figure~\ref{fig:shear}~(a) we show the probability distributions $P(G^m)$ of the local shear modulus $G^m$ at the indicated temperatures. In the figure, we also plot $P(G^m)$ at $T=0$, which is obtained with the harmonic approximation formulation~\cite{lutsko_1989,Mizuno2_2015}. We confirm that the distributions can be fitted by Gaussian functions at all values of $T$, and the average values are close to the $T=0$ value (red dashed line) and $T$-independent, as demonstrated in figure~\ref{fig:shear}~(b)~(red circles). By contrast, although for $T<10^{-1}$, the widths of the $P(G^m)$ stay close to the $T=0$ value, for higher temperatures, the distributions substantially broaden [figure~\ref{fig:shear}~(c), red circles]. Note that an analogous behaviour is hold by the distributions of the purely affine components, $G^m_\text A$ (blue squares).

In figure~\ref{fig:shear}~(d) (red circles), we plot our data in a different representation which will be useful below, and show the parameter $\gamma(T)=\hat{\rho} w^3 \delta G^2/G^2$, involving the ratio of $\delta G$ over $G$, normalized to the CG domain size $w$. This obviously follows the same behaviour as above, staying close to the $T=0$ value (red dashed line) for $T < 10^{-1}$ and next increasing rapidly when thermal fluctuations set in and induce substantial broadening. In the figure we also show the affine contribution to the shear moduli, $\gamma_\text A(T) = \hat{\rho} w^3 \delta G_\text A^2/G_\text A^2$ (blue squares), with again a $T$-dependence analogous to that of $\gamma(T)$. Note that $\gamma(T)$ ($\gamma_\text A(T)$) plays a crucial role in the HET~\cite{schirmacher_2006,schirmacher_2007,Schirmacher_2015,ferrante2013acoustic,marruzzo2013vibrational}, where the parameter $\delta\gamma=\gamma_\text c-\gamma$ (inset of figure~\ref{fig:shear}~(d), with $\gamma_\text c = 0.226$) controls the approach to the mechanical instability at $\gamma_\text c$ and induces the fractal frequency dependence of broadening, $\Gamma \propto \Omega^{3/2}$. We will address this point more in detail in section~\ref{subsect:discussion3}.
\begin{figure}[!b]
\centering
\includegraphics[width=0.87\textwidth]{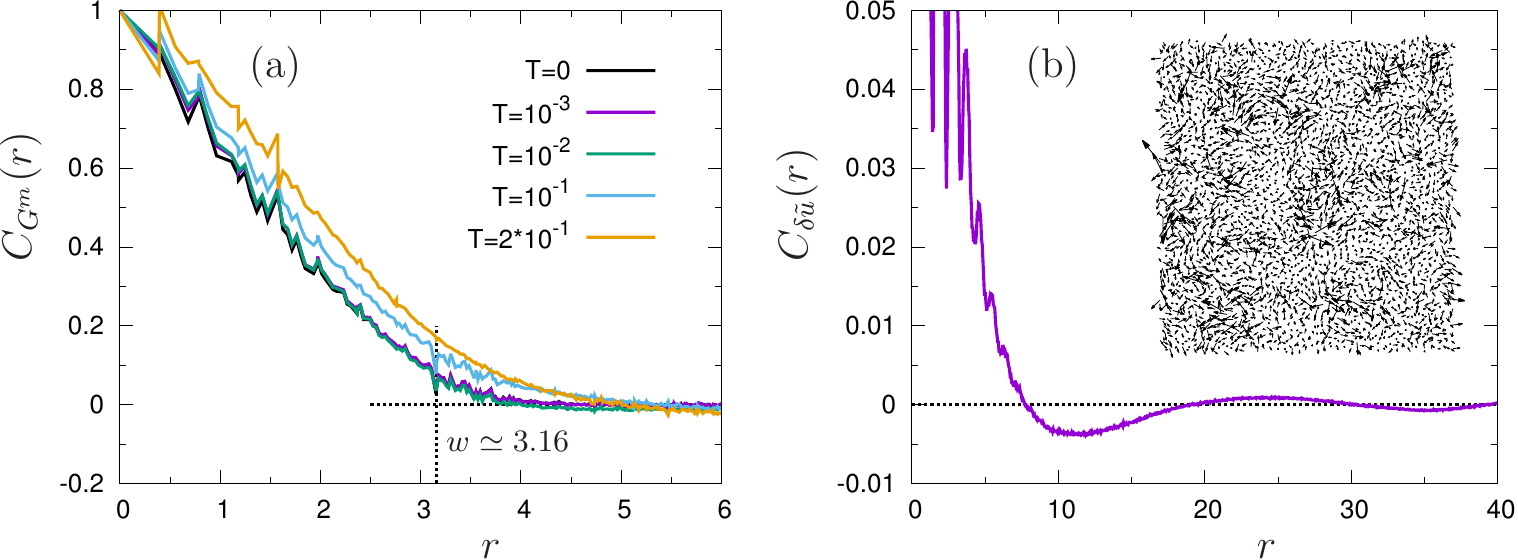}
\caption{(Colour online)
(a): Correlation functions $C_{G^m}(r)$ of the local shear modulus, equation~(\ref{eq:correlationf}), at the indicated $T$. The size of the coarse-graining domains, $w$, is indicated for reference. (b): The correlation function $C_{\delta \tilde{u}}(r)$ of the non-affine displacement field under a shear deformation, equation~(\ref{eq:correlationf2}). In the inset we represent by arrows the non-affine displacement field, $\delta \tilde{\mathbf{u}}_i$, within an arbitrary layer, for a system with $N=64000$.}
\label{fig:sheardis}
\end{figure}

{\bf Spatial correlations and the non-affine displacement field.} We have described above a clear crossover of broadening at $\Omega = \Omega_\text{co} \simeq 1$, where the scattering mechanism is modified from $\Gamma \propto \Omega^4$ to $\propto \Omega^2$. As we already mentioned, $\Omega_\text{co} \simeq 1$ is not far from the Ioffe-Regel frequency $\Omega_\text{IR} \simeq 2$~\cite{monaco2009anomalous}. As a consequence, at $\Omega_\text{co}$, the character of the vibrational excitations is substantially modified, crossing-over from quasi-plane waves to completely disordered envelopes of vibrational modes. We can associate to the crossover frequency a characteristic length-scale $\xi_\text{co}$ (or a wave number $q_\text{co} = 2\piup/\xi_\text{co}$), with $\xi_\text{co} = 2 \piup v_T/ \Omega_\text{co} \simeq 23$ ($q_\text{co} = \Omega_\text{co}/v_T \simeq 0.27$)~\cite{leonforte_2005}. We stress that both $\Omega_\text{co}$ and the length-scale $\xi_\text{co}$ are $T$-independent and related to the $T=0$ structural properties only. Here, we attempt to characterize $\xi_\text{co}$ in terms of the local shear modulus distributions.

In figure~\ref{fig:sheardis}~(a), we show the spatial correlation function of $G^m$, defined as
\begin{equation} 
\label{eq:correlationf}
C_{G^m}(r) = \frac{ \left< \left[G^m(\mathbf{r})-G \right]\left[G^m(\mathbf{0})-G \right] \right>_{\mathbf{0}} }{ \left<\left[G^m(\mathbf{0})-G \right]^2 \right>_{\mathbf{0}} }.
\end{equation}
Here, we have explicitly represented $G^m(\mathbf{r})$ as a function of the position $\mathbf{r}$, and the terms $[G^m(\mathbf{r})-G]$ are the fluctuations around the average (macroscopic) value, $G$. $\left< \right>_{\mathbf{0}}$ denotes the spatial average over $\mathbf{r}=\mathbf{0}$ and $r= \left| \mathbf{r} \right|$. Clearly, the $C_{G^m}(r)$ decay on length scales which, although mildly increasing with $T$ at the highest temperatures, always stay very close to the CG domains size $w$, implying no appreciable spatial correlations. We, therefore, conclude that $G^m$ fluctuates in space randomly, without any spatial correlation length, as also demonstrated for a-thermal jammed solids, both numerically~\cite{Mizuno2_2015} and in experiments~\cite{Zhang_2017}.

We now focus on the non-affine displacement field of particles. As reported in~\cite{DiDonna_2005}, randomly fluctuating local elastic moduli generate a non-affine displacement field~\footnote{More precisely, fluctuations of the affine elastic moduli, which are also randomly distributed in space~\cite{Mizuno_2013}, drive the non-affine displacement field.}, which exhibits a vortex-like structure. Interestingly, while the local elastic moduli show no correlations on length scales larger than that associated to the CG procedure, the non-affine displacement field is in contrast characterized by a typical length $\xi_\text{na}$. In~\cite{leonforte_2005}, it has been estimated that $\xi_\text{na}$ amounts to about $23$ particles size (diameter) for a 3-dimensional LJ system similar to the one studied here. It would therefore turn out that $\xi_\text{na}\simeq\xi_\text{co}$, extracted from the total transverse sound broadening, a point highlighted in the experimental work of~\cite{Baldi_2013}. 

To verify this point, we have extracted $\xi_\text{na}$ from the non-affine displacement field of our system, by following the procedure of~\cite{leonforte_2005}. This is based on a quasi-static shear deformation, where we first apply a small affine shear strain ($\gamma=5\times 10^{-4}$), and next relax the system to the closest local potential energy minimum. Throughout the relaxation process, we keep track of the total non-affine displacement, $\delta {\mathbf{u}}_i$, for all particles $i$.  As shown in~\cite{Maloney_2006}, additional elastic waves are excited during the non-affine relaxation, inducing system-spanning correlated motions of particles. It is, therefore, crucial to remove this contribution, by re-defining the non-affine displacement field as, $\delta \tilde{\mathbf{u}}_i = \delta {\mathbf{u}}_i - \sum_{k;\ \omega_k < \omega_0} \left[ \delta \mathbf{u}_i \cdot \mathbf{e}^k_i \right] \mathbf{e}^k_i$, with $\omega_k$ and $\mathbf{e}^k_i$ the eigenfrequencies and eigenvectors extracted by a normal mode analysis of the Hessian matrix~\cite{monaco2009anomalous}. Note that we set $\omega_0=1$ where the $T=0$ vDOS converges to the Debye limit~\cite{monaco2009anomalous}. 

We represent the $\delta \tilde{\mathbf{u}}_i$ with arrows in the inset of figure~\ref{fig:sheardis}~(b) where, by visual inspection, we indeed clearly recognize the vortex-like structure. To quantify the representative size of the latter, we have computed the spatial correlation function $C_{\delta \tilde{u}}(r)$,
\begin{equation} 
\label{eq:correlationf2}
C_{\delta \tilde{u}}(r) = \frac{ \left< \delta \tilde{\mathbf{u}}_i(\mathbf{r}_i)\cdot \delta \tilde{\mathbf{u}}_j(\mathbf{r}_j) \right> }{ \left< \delta \tilde{\mathbf{u}}_i(\mathbf{r}_i)\cdot \delta \tilde{\mathbf{u}}_i(\mathbf{r}_i) \right> },
\end{equation}
with $r=\left| \mathbf{r}_i - \mathbf{r}_j \right|$, and $\left< \right>$ denotes the average over all pairs of particles, $i$ and $j$. [Note that our $C_{\delta \tilde{u}}(r)$ does not show system-size effects, in contrast to the results of~\cite{Maloney_2006}, since we removed the elastic waves contributions described above]. The data of figure~\ref{fig:sheardis}~(b) show that $C_{\delta \tilde{u}}(r)$ decays on length scales $r=\xi_\text{na}\simeq 20$, consistently with~\cite{leonforte_2005} and very close to the $\xi_\text{co}\simeq 23$ associated to $\Omega_\text{co}$ as discussed above. 

This observation ultimately allows us to rationalize the value of the crossover frequency $\Omega_\text{co}$ in terms of the shear modulus heterogeneities as follows: {\em i)} The spatial fluctuations of the local shear modulus induce the non-affine character of the displacement field; {\em ii)} The latter is modulated in space, with correlations extending to $\lambda < \xi_\text{co}$, while for $\lambda > \xi_\text{co}$ correlations are lost; {\em iii)} As a consequence, the displacement field interacts with transverse modes differently for long and short length scales, corresponding to low ($\Omega < \Omega_\text{co}$) and high ($\Omega > \Omega_\text{co}$) frequencies, respectively; {\em iv)} These distinct effects induce the observed distinct scattering scalings, $\Gamma \propto \Omega^4$ and $\propto \Omega^2$, together with the associated cross-over at $\Omega_\text{co}$.
\subsection{Putting everything together: The heterogeneous elasticity theory}
\label{subsect:discussion3}
{\bf Sound broadening and the Boson peak.} In reference~\cite{monaco2009anomalous} we demonstrated a possible connection between the sound softening encoded in the pseudo-dispersion curves and the BP, by assuming $q$ as a good parameter for labelling vibrations in glasses and counting the number of acoustic modes in the low-$q$ region. This procedure reproduced the BP feature in the reduced vDOS~$\hat{g}(\omega)=g(\omega)/\omega^2$. Here, we adopt a different point of view, based on the HET and developed in reference~\cite{Marruzzo_2013} (the so-called generalized Debye model~\cite{Mizuno_2018}), which directly connects the transverse broadening $\Gamma$ to the BP features,
\begin{equation}
\hat{g}(\omega) = \frac{3}{\omega_\text D^3} +\frac{4}{\piup q_\text D^2 c^2(\omega)} \left[ \frac{\Gamma(\omega)}{\omega^2+\Gamma^2(\omega)} \right].
\label{eq:schirm}
\end{equation}
Here, $q_\text D=(6\piup^2\hat\rho)^{1/3} \simeq 3.92$, $c_\text D = \left[ \left(v_\text L^{-3}+2 v_\text T^{-3} \right)/3 \right]^{-1/3} \simeq 4.13$, and $\omega_\text D = q_\text D c_\text D \simeq 16.19$ are the the Debye wave-number, velocity and frequency, respectively, with $v_\text L = \sqrt{(K + 4G/3)/\rho} \simeq 8.71$ and $v_\text T = \sqrt{G/\rho} \simeq 3.65$ the longitudinal and transverse sound speeds. The term $3/\omega^3_\text D$ is the Debye limit. Hence, $\hat{g}(\omega)$ can be determined by equation~(\ref{eq:schirm}) if the sound speed, $c(\omega)$, and broadening, $\Gamma(\omega)$, are known, {\em including} all anharmonic effects at finite $T$. 

\begin{figure}[t]
\centering
\includegraphics[width=0.9\textwidth]{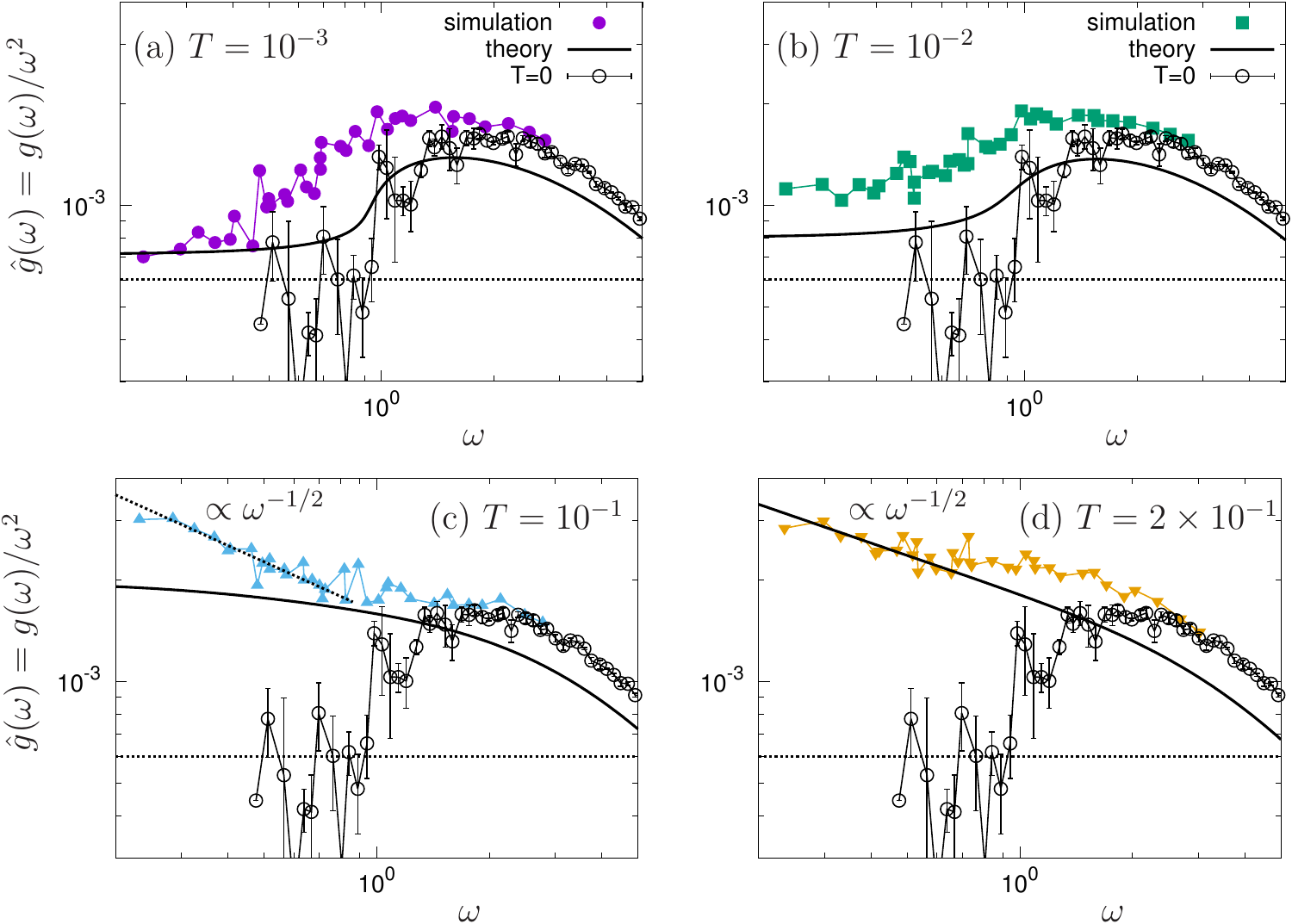}
\caption{(Colour online) (a)--(d): The reduced vDOS $\hat{g}(\omega)$ at the indicated temperatures. We compare the simulation data (closed symbols), obtained by equation~(\ref{eq:schirm}), and the HET predictions (solid lines). For reference, we also plot the values at $T=0$ (black open circles), obtained by diagonalizing the Hessian matrix~\cite{monaco2009anomalous}. The horizontal dashed lines indicate the Debye prediction, $3/\omega_\text D^3$.}
\label{fig:theory}
\end{figure}

In figure~\ref{fig:theory}~(a)--(d), we plot the right-hand side of equation~(\ref{eq:schirm}) at the indicated values of $T$ (closed symbols), together with $\hat{g}(\omega)$ at $T=0$, obtained by diagonalizing the Hessian matrix~\cite{monaco2009anomalous} (black open circles). The two sets of data in panel (a) are in nice qualitative agreement, with equation~(\ref{eq:schirm}) grasping the Debye limit, increasing quadratically in the Rayleigh range, and matching the BP intensity at $\omega_\text{BP}\simeq 2$. The situation is similar at $T=10^{-2}$ in (b), although anharmonicities already start to alter the small-$\omega$ behaviour, a modification which is completed at the two highest temperatures, $T=10^{-1}$ (c) and $2\times 10^{-1}$ (d). Now, the data decrease by increasing frequency, following an $\omega^{-1/2}$ power-law, before directly matching the BP intensity. The simultaneous presence of the $\omega^{-1/2}$ dependence and the BP corroborates the predictions of~\cite{marruzzo2013vibrational}, where the theory was modified to include an anharmonic scattering contribution~\cite{schirmacher2010sound,Tomaras_2010} responsible for the observed fractal behaviour. Similar data have been reported in the experimental work of~\cite{Baldi_2014}. 

{\bf Adjusting the HET to the numerical data.} Equation~(\ref{eq:schirm}) (and, therefore, the HET) indeed provides a convincing framework to describe our $T$-dependent vibrational data. It misses, however, an important aspect: it does {\em not} directly (and transparently) involve any quantity related to the local elastic properties, which ultimately control the observed physics. Note that in the theory these are {\em adjustable} parameters. In our simulations, by contrast, they can be measured (figure~\ref{fig:shear}) and are directly determined by the interaction potential and by the imposed external thermodynamic conditions. It, therefore, makes sense to {\em fit} the HET to our sound data, and next compare the obtained best-fit parameters to the elasticity measurements, providing a complete consistency check of the entire picture.

We have self-consistently solved the momentum conservation equation in the effective medium approximation~\cite{schirmacher_2006,schirmacher_2007,Schirmacher_2015}. In particular, we followed reference~\cite{marruzzo2013vibrational}, and solved the equations:
\begin{eqnarray}
&&\Sigma(\omega,T) = \Sigma_\text{harm}(\omega,T) + \Sigma_\text{anh}(\omega,T)\,,\nonumber \\
&&\Sigma_\text{harm}(\omega,T) = \gamma(T) \left(\frac{G_0}{\rho}\right)^2 \frac{1}{N} \sum_{|\mathbf{q}|<q_\text D} \left[ \chi_\text T(q,\omega) + \frac{2}{3} \chi_\text L(q,\omega) \right], \label{selfeq} \\
&&\chi_\text T(q,\omega) = \frac{q^2}{-\omega^2 + q^2 \left[ \frac{G_0}{\rho} - \Sigma(\omega,T) \right]}, \quad
\chi_\text L(q,\omega) = \frac{q^2}{-\omega^2 + q^2 \left[ \frac{K_0}{\rho} + \frac{4}{3}\left(\frac{G_0}{\rho} - \Sigma(\omega,T) \right) \right]}\,, \nonumber
\end{eqnarray}
where $\chi_\text{T,L}(q,\omega)$ are the transverse and longitudinal dynamical susceptibilities, respectively. The theory assumes an anharmonic coupling in the form of the Akhiezer-like scattering, amounting to an imaginary term $\Sigma_\text{anh}(\omega,T) \equiv \text{i} \nu \omega T$~\cite{schirmacher2010sound,Tomaras_2010}. Equations~(\ref{selfeq}) contain four free parameters: the shear ($G_0$) and bulk ($K_0$) moduli, the disorder parameter $\gamma=\hat{\rho} w^3 \delta G_0^2/G_0^2$, and the anharmonic coupling $\nu$. Since $\Sigma$ is directly related to sound velocity, broadening and vDOS~\cite{schirmacher_2006,schirmacher_2007,Schirmacher_2015}, as
\begin{equation}
\begin{aligned}
c(\omega) = \sqrt{\frac{G_0}{\rho} - \text{Re} \left[ \Sigma(
		\omega,T) \right]}\,\,, \quad \Gamma(\omega) = \frac{\omega \text{Im} \left[
	\Sigma(\omega,T) \right]}{c^2(\omega)}\,, \\
g(\omega) = \frac{2\omega}{3\piup} \text{Im} \left\{
\frac{1}{N} \sum_{|\mathbf{q}|<q_\text D} \frac{1}{q^2} \left[ 2 \chi_\text T(q,\omega)
+ \chi_\text L(q,\omega) \right] \right\},	
	\end{aligned} \nonumber	
	\end{equation}
 (where Re and Im denote the real and imaginary parts, respectively), we can tune the parameters to best reproduce those (MD) data via equations~(\ref{selfeq}). We have extracted the optimized ($T$-independent) values $G_0 = 25$, $K_0/G_0=2.4$, and $\nu \rho/G_0 = 3.0$~\footnote{For the case of $T=10^{-3}$, we set $\nu = 0$ because we did
 	not recognize any anharmonic contribution to the simulation data of
 	figure~\ref{fig:gamma}.}, while the variable strength of the elastic heterogeneity is encoded in the distance from the elastic instability, $\delta\gamma(T)=\gamma_\text c-\gamma(T)$, with $\gamma_\text c = 0.226$ [inset of figure~\ref{fig:shear}~(d)].

In figures~\ref{fig:gamma} and~\ref{fig:theory}, we plot by solid lines the predictions for $c(\Omega)$, $\Gamma(\Omega)$, and $\hat{g}(\omega)$ for all values of $T$, based on the procedure above. At the lowest $T=10^{-3}$, the theory acceptably reproduces the most distinctive features of the vibrational excitations, including the sound softening in $c(\Omega)$, the crossover in $\Gamma$ from the $\propto \Omega^4$ to the $\propto \Omega^2$ scaling, and the BP intensity and position in $\hat{g}(\omega)$. While in general the intensity of the broadening is considerably underestimated, the theory is capable of pin-pointing the crossover frequency, $\Omega_\text{co} \approx 1$, and thus the length scale, $\xi_\text{co} \approx 20$. Note that these values are directly determined by the parameter $\gamma$, since we do {\em not} explicitly insert those scales into the theory.
We also note that the effective medium theory for a-thermal systems~\cite{Wyart_2010,DeGiuli_2014} also strongly underestimates the attenuation, although correctly predicting the crossover frequency and length scales~\cite{Mizuno_2018}.

In addition, the HET also reasonably reproduces the effects due to anharmonicities at finite $T$ visible in our data, accounting for the disappearance of the sound softening in the $c(\Omega)$, the increase of anharmonic damping with temperature, and the fractal scaling laws $\propto \Omega^{3/2}$ in $\Gamma$ and $\propto \omega^{-1/2}$ in $\hat{g}(\omega)$ at the highest $T=2\times 10^{-1}$. In~\cite{mizuno2019unpub} we have been able to isolate the effects related to the anharmonic couplings and to demonstrate a clear correlation between the temperature variation of $\gamma(T)$ (related to the braoadening the distributions due  to an increasing fraction of negative shear stiffnesses on increasing $T$) and that (linear) of the strength of the Akhiezer-like mechanisms.
\begin{table*}[!b]
	\vspace{-4mm}
	\caption{
		Values of the HET parameters $G_0$, $K_0$, and $\gamma$, as extracted by fitting the theory to the MD data, or determined directly from the MD configurations by the computational methods summarized in~\cite{Mizuno_2013}. The values of $\gamma$ are estimated at $T=10^{-3}$. Note that all values of $\gamma$ determined from the MD configurations exceed the limit of validity of HET, $\gamma_\text c=0.226$.}
	\vspace{2ex}
\small
\centering
\begin{tabular}{c|c|c|c|c}
\hline
\hline
& HET fitting & Fully local approach & Affine components & Frozen matrix approach \\
\hline
$G_0$                & $25.0$ & $14.0$ & $35.6$ & $26.3$ \\
$K_0$                & $60.0$ & $59.7$ & $60.2$ & $60.0$ \\
$\gamma$ & $0.221$ ($\gamma_\text c=0.226$) & $3.49$ & $0.388$ & $0.820$  \\
\hline
\hline
\end{tabular}
\label{table:parameter}
\end{table*}

We can now provide additional insight by comparing the above best-fit values for $G_0$ and $K_0$ to those calculated from the MD configurations. In table~\ref{table:parameter} we report the values extracted by means of the indicated procedures~\cite{Mizuno_2013}, for a CG scale $w \simeq 3.16$. We have considered: {\em i)} the values shown in figure~\ref{fig:shear}~(b), (see section~\ref{sec:methods}); {\em ii)} the affine component values~[also shown in figure~\ref{fig:shear}~(b)], $G_\text A$ and $K_\text A$, where the non-affine contributions have been neglected; and {\em iii)} the values extracted via the {\em frozen matrix approach}~\cite{sollich_2009}. In the latter, the local elastic properties are determined by freezing all atoms located outside a target local region. The frozen part is next constrained to only relax affinely, amounting to a system restricted to only {\em localized} non-affine deformations~\cite{Mizuno_2013}. 

We immediately note that the optimized value for $K_0$ is matched by all methods, while completely disregarding the (negative) non-affine component, provides a $G_0$ which is strongly overestimated, as expected. On the contrary, we find that the fully local approach strongly underestimates $G_0$, which is not completely surprising either. Indeed, the HET does not explicitly treat the non-affine contributions to the elasticity on length scales of order $w$, which, therefore, must be absorbed in the values of $G_0$ and $\gamma$. This is in contrast to the fully local approach, which includes all deformations on length scales $w$, allowing one to dissipate more efficiently the non-affine motions comprised in the CG domain, and substantially decrease the shear modulus. As a consequence, it turns out that the frozen matrix procedure is the only one correctly grasping the mechanisms accounted for by the HET, providing a very similar value for $G_0$.

We conclude with an observation on our data for $\gamma(T)$ of figure~\ref{fig:shear}~(d), also reported in table~\ref{table:parameter}. These MD values exceed in all cases the limit of validity of the HET, $\gamma_\text c = 0.226$. This implies that if one simply inserts them in the HET, the theory would become unstable and fail badly. This limitation might originate from over-simplified assumptions, as the absence of spatial fluctuations of the local bulk modulus $K^m$, and could possibly be solved. Already at this level, however, the theory clearly provides a valuable insight, as demonstrated in the discussion above.

\section{Conclusions} 
\label{sect:conclusions} 
In this work we have elucidated the interplay of anharmonic couplings and the heterogeneous mechanical response at the nano-scale in determining transverse sound waves propagation in glasses. By modifying temperature, we have been able to control the relative strengths of these mechanisms. We have provided clear evidence that the sound softening encoded in the phase velocity $c(\Omega,T)$, which at very-low $T$ is completely determined by the disorder, is substantially decreased and eventually completely suppressed by anharmonicities. Even more intriguing, we have provided a complete characterization of the frequency dependence of the sound broadening, $\Gamma(\Omega,T)$, analyzing in depth the evolution of the sometimes elusive Rayleigh-like scattering, $\Gamma \sim \Omega^4$, in the intermediate $\Omega$-regime, and the disorder-controlled channel, $\Gamma \propto \Omega^2$, in the high-$\Omega$ region. We have also provided a complete description of the anharmonic channel, highlighting a fractal-like frequency scaling $\Gamma \propto \Omega^{3/2}$ at low frequencies.

In addition, we have demonstrated that the heterogeneous elasticity theory developed by W. Schirmacher and co-workers~\cite{schirmacher_2006,schirmacher_2007,Schirmacher_2015,marruzzo2013vibrational,schirmacher2010sound,Tomaras_2010}, is capable of acceptably reproducing the simulation data when fed with realistic nanoscale mechanics information. At the lowest $T$, the theory reproduces the sound softening as well as the crossover from $\Gamma \propto \Omega^4$ to $\propto \Omega^2$, both of which are disordered effects. The crossover frequency $\Omega_\text{co}\simeq 1$ and length $\xi_\text{co} = 2 \piup v_\text T/ \Omega_\text{co} \simeq 23$ are intimately related to the length $\xi_\text{na}$ associated to the non-affine displacement field, which is indeed induced by randomly fluctuating local elastic moduli. At the higher $T$, where the anharmonicities couple with the elastic heterogeneities, HET indeed reproduces a complicated sound propagation behavior, including disappearance of the sound softening, an increase of anharmonic damping with temperature, and the fractal-like scaling laws in $\Gamma(\Omega)$ and $\hat{g}(\omega)$. Remarkably, the theory predicts that the fractal scaling is related to the approach to the instability $\gamma_\text c$~\cite{marruzzo2013vibrational,ferrante2013acoustic}, a picture reasonably supported by our simulation data. Our results are an additional evidence that only developments which integrate local mechanical features of materials {\em and} a full treatment of the anharmonic couplings will be capable of providing the complete picture for sound waves propagation in disordered solids.
\vspace{-4mm}
\section*{Acknowledgements}
We thank Giancarlo Ruocco for useful discussions, and Walter Schirmacher for feedback in the early stage of this work. H. M. is supported by JSPS KAKENHI Grant Numbers 19K14670 and the Asahi Glass Foundation. S. M. is supported by ANR-18-CE30-0019 (HEATFLOW).
\bibliographystyle{cmpj}

\newpage
\ukrainianpart

\title{Вплив еластичних неоднорідностей на поширення коливань при скінчених температурах у скловидних системах}
\author{Г. Мізуно\refaddr{label1} , С. Мосса\refaddr{label2,label3}}
\addresses{
	\addr{label1} Школа мистецтв та наук, Університет Токіо, Токіо 153-8902, Японія
	\addr{label2} Університет Гренобля Альп, CEA, CNRS, IRIG, SyMMES, F-38000 Гренобль, Франція
	\addr{label3} Інститут Лауе-Ланжевена, BP 156, F-38042 Гренобль, Франція
}

\makeukrtitle

\begin{abstract}
Деякі аспекти поширення звукових хвиль через невпорядковані тверді тіла все ще є неясними. Недавня робота
описувала особливість невпорядкованих твердих тіл, яка впливає на коливні збудження на проміжних масштабах,
- еластичну неоднорідність. Було показано, що поширення звукових хвиль зазнає сильного впливу від неоднорідних
механічних особливостей матеріалів, які додаються до стандартних ангармонічних взаємодій, результуючи в 
надзвичайно складних явищах переносу при скінчених температурах. Тут ми досліджуємо ці проблеми, маючи 
випадок простої атомарної склоформуючої системи, комп'ютерним моделюванням методом молекулярної динаміки. 
Зокрема, ми фокусуємся на поперечних компонентах коливних збуджень через динамічні структурні фактори, і 
характеризуємо температурну залежність дисперсії звуку та загасання в широкій області частот. Ми подаємо
повну картину того, як еластична неоднорідність визначає перенос коливних збуджень, також на основі 
прямого порівняння чисельних результатів з передбаченням теорії неоднорідної еластичності.

\keywords квазічастинки і колективні збудження, аморфні матеріали, молекулярна динаміка

\end{abstract}

\end{document}